\documentclass[prl,twocolumn,preprintnumbers,superscriptaddress,showpacs]{revtex4-2}
\usepackage[colorlinks=true, pdfstartview=FitV, linkcolor=red, citecolor=blue, urlcolor=blue, pdftitle={},pdfsubject={}, pdfkeywords={}]{hyperref}

\newcommand\sect[1]{{\it #1.}---}

\usepackage{stmaryrd}
\usepackage{comment}
\usepackage{color}
\usepackage{amsmath}
\usepackage{amssymb}
\usepackage{amsthm}
\usepackage{mathrsfs}
\usepackage{graphicx}
\usepackage{marvosym}
\usepackage{amsmath,bm}
\usepackage{array}
\usepackage{latexsym}
\usepackage{enumerate}
\usepackage{slashed}
\usepackage{hyperref}
\usepackage{color}

\allowdisplaybreaks[1]

\usepackage[normalem]{ulem}

\begin{document}
\title{Effective Chiral Magnetic Effect from Neutrino Radiation}
\author{Naoki Yamamoto}
\affiliation{
Department of Physics, Keio University, Yokohama 223-8522, Japan.\\
}
\author{Di-Lun Yang}
\affiliation{
Institute of Physics, Academia Sinica, Taipei 11529, Taiwan.\\
}
\begin{abstract}
We develop an approach to chiral kinetic theories for electrons close to equilibrium and neutrinos away from equilibrium based on a systematic power counting scheme for different timescales of electromagnetic and weak interactions. Under this framework, we derive electric and energy currents along magnetic fields induced by neutrino radiation in general nonequilibrium states. This may be regarded as an effective chiral magnetic effect (CME), which is present without a chiral chemical potential, unlike the conventional CME. We also consider the so-called gain region of core-collapse supernovae as an example and find that the effective CME enhanced by persistent neutrino emission in time is sufficiently large to lead to the inverse cascade of magnetic and fluid kinetic energies and observed magnitudes of pulsar kicks. Our framework may also be applicable to other dense-matter systems involving nonequilibrium neutrinos.
\end{abstract}

\maketitle

\sect{Introduction}%
In chiral matter composed of approximately massless fermions with chiral imbalance, an electric current is induced by magnetic fields. This chiral magnetic effect (CME)~\cite{Vilenkin:1980fu,Nielsen:1983rb,Alekseev:1998ds,Fukushima:2008xe} has been widely studied in a variety of physical systems, such as heavy ion collisions~\cite{Kharzeev:2015znc}, early Universe~\cite{Kamada:2022nyt}, compact stars~\cite{Kamada:2022nyt}, and Dirac-Weyl semimetals~\cite{Armitage:2017cjs,Gorbar2021}. Moreover, the presence of electric currents from CME results in unstable modes for dynamically growing magnetic fields, known as chiral plasma instability (CPI)~\cite{Joyce:1997uy,Akamatsu:2013pjd}, which has multiple applications particularly in cosmology and astrophysics~\cite{Kamada:2022nyt}.   

On the other hand, the magnitude of chiral imbalance is expected to be small in most physical systems due to the absence of intrinsic parity violation. The exception is systems involving the weak interaction that globally violates parity symmetry. One example is the electron capture and its inverse process, ${\rm e}^{-}_{\rm L}+{\rm p} \leftrightarrow \nu_{\rm L}+{\rm n}$, in core-collapse supernovae (CCSN), where chiral imbalance of leptons could be generated~\cite{Charbonneau:2009ax,Akamatsu:2013pjd,Ohnishi:2014uea,Kaminski:2014jda,Yamamoto:2015gzz}. Accordingly, the CME and CPI may be triggered, which affect the dynamics of the matter sector composed of electrons and nucleons. Nevertheless, such chiral imbalance in the electron sector could be washed out by chirality flipping due to a small yet nonzero electron mass in thermal equilibrium~\cite{Grabowska:2014efa,Kaplan:2016drz}. 

This scenario is expected to be modified when neutrinos are out of equilibrium. In fact, recent studies based on the newly developed chiral radiation transport theory for neutrinos that includes the chiral effects~\cite{Yamamoto:2020zrs} suggest the presence of electric and energy currents of matter along magnetic fields driven by the backreaction of neutrinos slightly away from equilibrium~\cite{Yamamoto:2021hjs,Matsumoto:2022lyb}. This may be regarded as an effective CME, which is present even without chiral imbalance. Although these chiral effects from neutrino radiation not close to equilibrium should be prominent in practical applications (e.g., outside the core of CCSN), such a derivation has been lacking to date.     

In this Letter, we, for the first time, derive the effective CME of electrons sourced by neutrino radiation out of equilibrium. For this purpose, we use the chiral kinetic theories~\cite{Son:2012wh,Stephanov:2012ki,Son:2012zy,Chen:2012ca,Manuel:2013zaa,Manuel:2014dza,Chen:2015gta,Hidaka:2016yjf,Hidaka:2017auj,Huang:2018wdl} incorporating the chiral effects for ultrarelativistic electrons close to equilibrium and neutrinos away from equilibrium, with the collision term of the neutrino absorption on nucleons and its inverse process. We develop a systematic power counting scheme for different timescales of electromagnetic and weak interactions. 
Even for the conventional radiation hydrodynamics for neutrinos and matter, a systematic power counting scheme in the spirit of the low-energy effective theory has not been explicitly provided in the literature, to the best of our knowledge. Our scheme not only provides a theoretical foundation as such, but also enables us to obtain the effective CME from radiation of neutrinos in generic nonequilibrium states, allowing for broader applications than previously.

We also show that this effective CME due to purely nonequilibrium interaction is enhanced by persistent neutrino emission in time and provides a dominant contribution compared with the previous results~\cite{Yamamoto:2021hjs,Matsumoto:2022lyb} that are suppressed in the nonrelativistic expansion for nucleons. Given a quasithermal distribution function of nonequilibrium neutrinos in the so-called gain region of CCSN, we further estimate the numerical values of the effective chiral magnetic conductivity for the electric and energy currents. We find that they can reach sufficiently large magnitudes to lead to the inverse cascade of magnetic and fluid kinetic energies and observed magnitudes of pulsar kicks. 

Throughout this Letter, we use the mostly minus signature of the Minkowski metric $\eta^{\mu\nu}$ and the completely antisymmetric tensor $ \epsilon^{\mu\nu\rho\lambda} $ with $ \epsilon^{0123} = 1$. We introduce the shorthand notations $A^{(\mu}B^{\nu)}\equiv A^{\mu}B^{\nu}+A^{\nu}B^{\mu}$ and $A^{[\mu}B^{\nu]}\equiv A^{\mu}B^{\nu}-A^{\nu}B^{\mu}$ and define $\tilde{F}^{\mu\nu}\equiv\epsilon^{\mu\nu\alpha\beta}F_{\alpha\beta}/2$ with $F^{\mu\nu}$ being the electromagnetic field strength. We also set $c=k_{\rm B}=1$, but keep $\hbar$ to show an expansion of the quantum corrections unless stated otherwise.

\sect{Chiral kinetic equations for electrons near equilibrium}%
The chiral kinetic equation for electrons with chirality $\chi = \pm 1$ (denoted by the subscript $\chi = {\rm R, L}$) takes the form~\cite{Yamamoto:2020zrs,Kamada:2022nyt}  
\begin{eqnarray}
\Box_q f_{\chi}^{(\rm e)}=(1-f_{\chi}^{(\rm e)}) \Gamma^{<}_{\chi} - f_{\chi}^{(\rm e)} \Gamma^{>}_{\chi}\,,
\end{eqnarray}
accompanied by the on-shell condition,
\begin{eqnarray}
	q^2=-\chi\hbar S^{\alpha\beta}_{q}eF_{\alpha\beta}.
\end{eqnarray}
Here, $f_{\chi}^{(\rm e)}$ is the distribution function of electrons, the operator $\Box_q$ is defined as
\begin{eqnarray}
\Box_qf_{\chi}^{(\rm e)}=\left(q^{\mu}+\chi\hbar\frac{S^{\mu\nu}_{q}eF_{\mu\rho}n^{\rho}}{q\cdot n}\,\right) \Delta_{\mu} f_{\chi}^{(\rm e)},
\end{eqnarray}
with $\Delta_{\mu}=D_{\mu}+eF_{\lambda\mu}\partial_{q}^{\lambda}$ and $D_{\mu}=\nabla_{\mu}-\Gamma^{\lambda}_{\mu\rho}q^{\rho}\partial_{q\lambda}$, $S^{\mu\nu}_{q}=\epsilon^{\mu\nu\alpha\beta}q_{\alpha}n_{\beta}/(2q\cdot n)$ is the spin tensor associated with a frame vector $n^{\mu}=(1,\,{\bm 0})$~\footnote{The choice of a frame vector is similar to the gauge choice, which does not affect the final results of physical observables~\cite{Chen:2015gta,Hidaka:2017auj}.}, and $e$ is the electron charge. Also, $\Gamma^{\lessgtr}_{\chi} = q \cdot \Sigma^{\lessgtr}_{\chi}$
are proportional to the emission and absorption rates of electrons with $\Sigma^{\lessgtr}_{\nu}$ being the lesser and greater self energy that can be obtained from the underlying scattering processes. In most cases, $\Sigma^{\lessgtr}_{\chi,\mu}\propto n_{\mu}, q_{\mu}$ up to $\mathcal{O}(\hbar^0)$, since further anisotropic contributions stem from gradient terms of $\mathcal{O}(\hbar)$, and hence, $\hbar S^{\mu\nu}_q\Sigma^{\lessgtr}_{\chi,\mu} = \mathcal{O}(\hbar^2)$ are suppressed above; for the complete expression with these terms, see Refs.~\cite{Yamamoto:2020zrs,Kamada:2022nyt}.

From now on, we will mostly focus on left-handed electrons. We can further decompose the collision term into two parts, $\Gamma^{\lessgtr}_{\rm L}=\Gamma^{\lessgtr}_{\rm EM}+\Gamma^{\lessgtr}_{\rm W}$, where the subscripts ``$\rm EM$" and ``$\rm W$" represent the electromagnetic and weak interactions, respectively. When electrons are near thermal equilibrium, we can decompose $f_{\chi}^{(\rm e)}=\bar{f}_{\chi}^{(\rm e)}+\delta f_{\chi}^{(\rm e)}$ with $|\delta f_{\chi}^{(\rm e)}|\ll |\bar{f}_{\chi}^{(\rm e)}|$, where $\bar{O}$ represents a physical object $O$ in thermal equilibrium and $\delta O$ corresponds to the small fluctuation. 
In such a case, we may also approximate
\begin{eqnarray}
	\Gamma^{\lessgtr}_{\rm L}\approx\bar{\Gamma}^{\lessgtr}_{\rm EM}+\delta \Gamma^{\lessgtr}_{\rm EM}+\Gamma^{\lessgtr}_{\rm W},
\end{eqnarray}
where $\delta\Gamma^{\lessgtr}_{\rm EM}$ are functions of $\delta f_{\chi}^{(\rm e)}$ up to linear in $\delta f_{\chi}^{(\rm e)}$.
Detailed balance in thermal equilibrium entails 
\begin{eqnarray}\label{eq:eql_cond_EM}
	(1-\bar{f}_{\rm L}^{(\rm e)}) \bar{\Gamma}^{<}_{\rm EM}= \bar{f}_{\rm L}^{(\rm e)}\bar{\Gamma}^{>}_{\rm EM},
\end{eqnarray}
and the collision term becomes
\begin{eqnarray}\nonumber\label{eq:collision_term}
	&&(1-f_{\rm L}^{(\rm e)}) \Gamma^{<}_{\rm L} - f_{\rm L}^{(\rm e)} \Gamma^{>}_{\rm L}
	\\\nonumber
	&&\approx (1-\bar{f}_{\rm L}^{(\rm e)}) (\delta \Gamma^{<}_{\rm EM}+\Gamma^{<}_{\rm W})-\bar{f}_{\rm L}^{(\rm e)}(\delta\Gamma^{>}_{\rm EM}+\Gamma^{>}_{\rm W})
	\\
	&&\quad-\delta f_{\rm L}^{(\rm e)}(\bar{\Gamma}^{>}_{\rm EM}+\bar{\Gamma}^{<}_{\rm EM})
\end{eqnarray}
up to the terms linear to nonequilibrium fluctuations [i.e., $\mathcal{O}(|\delta \Gamma^{\lessgtr}_{\rm EM}|)$], where we have also neglected the subleading contributions $\Gamma^{\lessgtr}_{\rm W}\delta f_{\rm L}^{(\rm e)}$.

Given the above approximations, one may recast part of the collision term associated with electromagnetic interaction into the form of a relaxation time:
\begin{eqnarray}\nonumber
	&&(1-\bar{f}_{\rm L}^{(\rm e)})\delta \Gamma^{<}_{\rm EM}-\bar{f}_{\rm L}^{(\rm e)}\delta\Gamma^{>}_{\rm EM}-\delta f_{\rm L}^{(\rm e)}(\bar{\Gamma}^{>}_{\rm EM}+\bar{\Gamma}^{<}_{\rm EM})
	\\
	&&\approx -q\cdot n\hat{\tau}_{\rm EM}^{-1}\delta f_{\rm L}^{(\rm e)}.
\end{eqnarray}
Note that $\hat{\tau}_{\rm EM}^{-1}$ can be an operator acting on $\delta f_{\rm L}^{(\rm e)}$ in general, while its exact form is not of our interest here. Then the chiral kinetic equation for left-handed electrons takes the form
\begin{eqnarray}\label{eq:eCKT_relax}
	\Box_qf_{\rm L}^{(\rm e)}
	\approx  -q\cdot n\hat{\tau}_{\rm EM}^{-1}\delta f_{\rm L}^{(\rm e)}-F_{\rm W},
\end{eqnarray}
where 
\begin{eqnarray}
	F_{\rm W}=\bar{f}_{\rm L}^{(\rm e)}\Gamma^{>}_{\rm W}-(1-\bar{f}_{\rm L}^{(\rm e)}) \Gamma^{<}_{\rm W}
\end{eqnarray}
is responsible for the backreaction of neutrino radiation upon the matter sector. In the above expression, we neglect nonequilibrium fluctuations of nucleons that are expected to be suppressed by large masses. 
We also omit the collision term $\propto \delta f^{({\rm e})}_{\rm R}$ related to chirality flipping via the electromagnetic interaction. As will be manifested later, such a term does not affect the nonequilibrium transport of electrons due to neutrino radiation as our primary concern, similar to the term $q\cdot n\hat{\tau}_{\rm EM}^{-1}\delta f_{\rm L}^{(\rm e)}$. 
Below we will consider the situation where $\bar f_{\rm L}^{(\rm e)} = \bar f_{\rm R}^{(\rm e)}$, which corresponds to the vanishing chiral chemical potential of electrons, $\mu_5 = 0$.

For the total nonequilibrium corrections on the electric current and energy-momentum tensor, which can be calculated via Wigner functions, we have to include the contributions from both right- and left-handed electrons. Recall that the Wigner functions for right- and left-handed electrons take the form~\cite{Hidaka:2016yjf,Liu:2018xip,Yamamoto:2020zrs}
\begin{eqnarray}\label{WF_L_full}\nonumber
	\mathcal{W}^{< \mu}_{\chi}&=&2\pi\Big[\delta(q^2)\big(q^{\mu}+ \chi \hbar S^{\mu\nu}_{q} \Delta_{\nu}\big) 
	\\
	&&+ \chi \hbar e \tilde{F}^{\mu\nu}q_{\nu}\delta'(q^2)\Big]f_{\chi}^{(\rm e)}\,,
\end{eqnarray} 
where we suppressed the terms $\hbar S^{\mu\nu}_q\Sigma^{\lessgtr}_{\chi,\mu} = \mathcal{O}(\hbar^2)$ as above.
The electric current and energy-momentum tensor for electrons are given by
\begin{eqnarray}
	\label{j}
	j^{\mu}_{\chi}&=&2e\int\frac{{\rm d}^4q}{(2\pi)^4} \mathcal{W}^{< \mu}_{\chi}\,,
	\\
	T^{\mu\nu}_{\chi}&=&\int\frac{{\rm d}^4q}{(2\pi)^4} \mathcal{W}^{< (\mu}_{\chi} q^{\nu)}\,.
\end{eqnarray}
Note that $F_{\rm W}$ can generate finite $\delta f_{\rm L}^{(\rm e)} - \delta f_{\rm R}^{(\rm e)}$ even when $\bar f_{\rm L}^{(\rm e)} - \bar f_{\rm R}^{(\rm e)} = 0$ (or $\mu_5=0$).

\sect{Collision term for neutrino absorption on nucleons}%
To obtain an explicit form of $F_{\rm W}$, we shall focus on the neutrino absorption on nucleons and its inverse process, where electrons and nucleons are approximately in thermal equilibrium but neutrinos are not. Following construction of the collision term for the chiral radiation transport equation of left-handed neutrinos in Ref.~\cite{Yamamoto:2020zrs}, one can analogously derive $\Sigma^{\lessgtr}_{{\rm W}\mu}$ and $F_{\rm W}$, while the contributions from the electromagnetic interaction on self-energies can be calculated independently.

Using the nonrelativistic approximation for nucleons, ignoring the mass difference between protons and neutrons, and implementing the isoenergetic approximation, we find
\begin{eqnarray}\label{eq:FW_general}\nonumber
	F_{\rm W}&\approx &\frac{(q\cdot u)^3}{\pi}\big(g_{\rm V}^2+3g_{\rm A}^2\big){G}_{\rm F}^2 (n_{\rm p}-n_{\rm n})\bigg[
	\frac{\bar{f}^{(\rm e)}(1-f^{(\nu)})}{1-{\rm e}^{\beta(\mu_{\rm n}-\mu_{{\rm p}})}}
	\\
	&&
	+\frac{(1-\bar{f}^{(\rm e)})f^{(\nu)}}{1-{\rm e}^{\beta(\mu_{\rm p}-\mu_{{\rm n}})}}\bigg],
\end{eqnarray}
where we took $\bar{f}_{{\rm L}}^{(\rm e)}=\bar{f}_{{\rm R}}^{(\rm e)}=\bar{f}^{(\rm e)}$ with $\bar{f}^{(\rm e)}\equiv 1/({\rm e}^{\beta(q\cdot u-\mu_{\rm e})}+1)$. Here, $u^{\mu}$ is the fluid four-velocity, $n_{i}$ and $\mu_{i}$ for $i=\rm n,\,\rm p,\, \rm e$ denote the number densities and chemical potentials of corresponding particles, respectively, $f^{(\nu)}$ is the neutrino distribution function, and $\beta=1/T$ with $T$ being temperature. Also, $g_{\rm V}$ and $g_{\rm A}$ are vector and axial-vector couplings in Fermi's effective theory for weak interaction and $G_{\rm F}$ is the Fermi constant. Generically, $f^{(\nu)}$ has to be obtained by solving the chiral transport equation for neutrinos. It is easy to check that $F_{\rm W}=0$ in $\beta$ equilibrium.

\sect{Systematic power counting and effective CME}%
Since the collision term of the kinetic equation for left-handed electrons incorporates the interactions with distinct timescales, we have to modify the standard relaxation-time approximation to evaluate the nonequilibrium fluctuations.   
To have a description of kinetic equations consistent with radiation hydrodynamics, we postulate $f_{\rm L}^{(\rm e)}=\bar{f}_{\rm L}^{(\rm e)}+\delta f_{\rm L,EM}^{(\rm e)}+\delta f_{\rm L,W}^{(\rm e)}$, where $\delta f_{\rm L,EM}^{(\rm e)} \sim \hat{\tau}_{\rm EM}/L \ll 1$ as the gradient expansion with $L$ being the system size, while the expansion for $\delta f_{\rm L,W}^{(\rm e)}$ is based on the small expansion parameter related to the weak coupling, $\delta f_{\rm L,W}^{(\rm e)} \sim \epsilon^4G_{\rm F}^2 \ll 1$ with $\epsilon$ being the typical energy scale in the system. This power counting scheme makes it feasible to disentangle the backreaction on the matter sector due to the weak interaction systematically.

It then follows that 
\begin{eqnarray}\label{eq:deltafEM_KE}
	\Box_q\bar{f}_{\rm L}^{(\rm e)}\approx-q\cdot n\hat{\tau}_{\rm EM}^{-1}\delta f_{\rm L,EM}^{(\rm e)}
\end{eqnarray}
and
\begin{eqnarray}\label{eq:deltafW_KE}
	\Box_q\delta f_{\rm W}^{(\rm e)}\approx (1-\bar{f}_{\rm L}^{(\rm e)}) \Gamma^{<}_{\rm W}-\bar{f}_{\rm L}^{(\rm e)}\Gamma^{>}_{\rm W}=-F_{\rm W}.
\end{eqnarray}
Note that the term $-q\cdot n\hat{\tau}_{\rm EM}^{-1}\delta f_{\rm W}^{(\rm e)}\sim \epsilon^4G_{\rm F}^2 e^4$ is subleading in both Eqs.~(\ref{eq:deltafEM_KE}) and (\ref{eq:deltafW_KE}), and provides the higher-order corrections to the transport coefficients of the matter sector, and hence, it is dropped.
We accordingly find
\begin{eqnarray}
	\delta f_{\rm L,EM}^{(\rm e)} \approx-\hat{\tau}_{\rm EM}(q\cdot n)^{-1}\Box_q\bar{f}_{\rm L}^{(\rm e)}.
\end{eqnarray}
When considering chirality flipping, one shall find $\delta f_{\rm L,EM}^{(\rm e)}$ and $\delta f_{\rm R,EM}^{(\rm e)}$ in terms of the linear combination of $\Box_q\bar{f}_{\rm L}^{(\rm e)}$ and $\Box_q\bar{f}_{\rm R}^{(\rm e)}$ based on the coupled chiral kinetic equations.

On the other hand, $\delta f_{\rm W}^{(\rm e)}$ has to be solved from Eq.~(\ref{eq:deltafW_KE}) separately. For simplicity, we shall work in the Minkowski spacetime such that $D_{\mu}=\partial_{\mu}$. As we will eventually be interested in the regime of sufficiently large scale where the electric field $E_{\mu}=F_{\mu\rho}n^{\rho}$ is screened, let us  focus on momentum anisotropy of neutrinos induced by the magnetic field $B^{\mu}=\tilde{F}^{\mu\nu}n_{\nu}$. Then, the left-hand side of Eq.~(\ref{eq:deltafW_KE}) reduces to $q^{\mu}(\partial_{\mu}+\epsilon_{\mu\rho\alpha\beta}eB^{\alpha}n^{\beta}\partial^{\rho}_{q})\delta f_{\rm W}^{(\rm e)}$. For convenience, we introduce the shorthand notation for the spatial component, $\bar V^{\mu} \equiv \Theta^{\mu\nu}V_{\nu}$, of an arbitrary vector $V^{\mu}$ using the projection operator $\Theta^{\mu\nu}\equiv \eta^{\mu\nu}-n^{\mu}n^{\nu}$. Also, we always work in the fluid rest frame such that $u^{\mu} = n^{\mu}$. In the present setup, we expect $\partial^{\rho}_{q}\delta f_{\rm W}^{(\rm e)}\propto u^{\rho},\, q^{\rho},\, B^{\rho}$, so the Lorentz-force term  in the kinetic theory, $\epsilon_{\mu\rho\alpha\beta}q^{\mu} eB^{\alpha}n^{\beta}\partial^{\rho}_{q}\delta f_{\rm W}^{(\rm e)}$, identically vanishes. Consequently, Eq.~(\ref{eq:deltafW_KE}) reduces to
\begin{eqnarray}
	q\cdot \partial\delta f_{\rm W}^{(\rm e)}\approx -F_{\rm W}.
\end{eqnarray}

For a generic differential equation,
\begin{eqnarray}
	q\cdot\partial f(q, x)=G(q,x),
\end{eqnarray}
with $G(q,x)$ being an arbitrary function, the retarded solution of $f(q, x)$ is given by using the method of characteristics as (see also Ref.~\cite{Yang:2021fea})
\begin{eqnarray}
	f(q,x)=\frac{1}{q_0}\int^{x_0}_{-\infty} {\rm d}x'_0 G(q,x')|_{\text{c}}\,,
\end{eqnarray}
where $|_{\text{c}}=\{x^{\prime\mu}_{\perp}=x^{\mu}_{\perp},x^{\prime\mu}_{\parallel}=x^{\mu}_{\parallel} - \bar q^{\mu} (x_0-x'_0)/q_0\}$. 
Here, $V^{\mu}_{\parallel} \equiv (V \cdot \bar q) \bar q^{\mu}/ ({\bar q} \cdot {\bar q})$ and $V^{\mu}_{\perp} \equiv \bar V^{\mu} - V^{\mu}_{\parallel}$ represent the parallel and perpendicular components of a vector $V^{\mu}$ with respect to $\bar q^{\mu}$, respectively. From the useful equation above, we obtain
\begin{align}
	\delta f_{\rm W}^{(\rm e)}(q,x)=-\frac{1}{q_0}\int^{x_0}_{0} {\rm d}x'_0 F_{\rm W}(q,x')|_{\text{c}}\,,
\end{align}
where the explicit expression of $F_{\rm W}$ is shown in Eq.~(\ref{eq:FW_general}).
Note that here
$q^{\mu}\epsilon_{\mu\rho\alpha\beta}B^{\alpha}n^{\beta}\partial^{\rho}_{q}\delta f_{\rm W}^{(\rm e)}=0$ is satisfied. 

\begin{figure}
	\begin{center}
		\includegraphics[width=0.6\hsize]{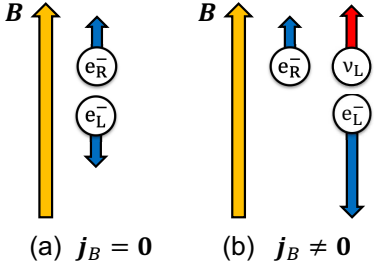}
	\end{center}
	\vspace{-0.5cm}
	\caption{The schematic figure of the effective CME from neutrino radiation.}
	\label{eCME_figure}
\end{figure}

In this case, the (spatial) electric and energy currents of electrons induced by the backreaction of neutrino radiation read
\begin{align}
	j_{B}^{\mu}
	\approx \hbar e^2 \int\frac{{\rm d}^4q}{(2\pi)^3}\frac{\delta(q^2)}{q_0} \left(B^{\mu}q\cdot\partial_{q}-q\cdot B\partial^{\mu}_{\bar q}\right)\delta f_{\rm W}^{(\rm e)}, \\
	T_{B}^{\mu 0}
	\approx \frac{\hbar e}{2} \int\frac{{\rm d}^4q}{(2\pi)^3}\delta(q^2)\left(B^{\mu}q\cdot\partial_{q}-q\cdot B\partial^{\mu}_{\bar q}\right)\delta f_{\rm W}^{(\rm e)}
\end{align}
for $q_0\geq 0$. Here, we used Eq.~(\ref{WF_L_full}) and performed integration by parts on the last term to derive the above expressions. 
Note that although half of the left-handed electrons that receive the backreaction from neutrinos are converted to right-handed ones at the timescale $\tau \gg \hat \tau_{\rm EM}$ due to chirality flipping, this does not affect the above macroscopic currents at $\mu_5 = 0$.
As schematically shown in Fig.~\ref{eCME_figure}, while the CME for right- and left-handed electrons in equilibrium cancel each other [Fig.~\ref{eCME_figure}(a)], the neutrino radiation triggers the effective CME with a nonvanishing electric current [Fig.~\ref{eCME_figure}(b)].  

As a particular limit of this formulation, we can also consider the case where neutrinos are close to equilibrium. In this case, we can take $f^{(\nu)}= \bar{f}^{(\nu)}+\delta f^{(\nu)}$ and rewrite $F_{\rm W}$ into the form of the relaxation time approximation,
\begin{eqnarray}
	F_{\rm W}\approx q\cdot n\tau^{-1}_{\rm W}\delta f^{(\nu)}.
\end{eqnarray}
Combining with the chiral transport equation for neutrinos near equilibrium~\cite{Yamamoto:2021hjs}, the kinetic equation of left-handed electrons implies
\begin{eqnarray}
\delta f_{\rm L,W}^{(\rm e)}\approx -\delta f^{(\nu)}=-\tau_{\rm W}(q\cdot n)^{-1} q\cdot D\bar{f}^{(\nu)}.
\end{eqnarray} 
Note that $\delta f_{\rm L,W}^{(\rm e)}\gg \delta f_{\rm L,EM}^{(\rm e)}$ since $\tau_{\rm W}\gg \hat{\tau}_{\rm EM}$. This is consistent with the previous derivation based on the momentum conservation in Ref.~\cite{Matsumoto:2022lyb}.

\sect{Numerical estimates in CCSN}%
Let us now estimate the magnitude of the effective CME in CCSN as an example. For this purpose, we adopt the analytic form of the neutrino distribution function introduced in Ref.~\cite{Keil:2002in}, 
\begin{eqnarray}
	f^{(\nu)}(q_0)=\left(\frac{q_0}{\bar{\epsilon}}\right)^{\alpha}{\rm e}^{-(\alpha+1)q_0/\bar{\epsilon}}\,,
\end{eqnarray} 
where $\alpha$ is a numerical parameter describing spectral pinching and $\bar{\epsilon}$ is the average energy.
In such a case, $F_{\rm W}$ only depends on $q_0$, and hence, 
\begin{eqnarray}
	\delta f_{\rm W}^{(\rm e)}(q,x)\approx -\frac{x_0}{q_0}F_{\rm W}(q_0)\,,
\end{eqnarray}
which takes a secular form with respect to the elapsed time $x_0$.
One then finds
\begin{eqnarray}\nonumber
	j_{B}^{\mu}
	&\approx &
	\frac{\hbar e^2}{4\pi^2} x_0 B^{\mu} \int^{\infty}_0 \frac{{\rm d}|\bm q|}{|\bm q|}F_{\rm W}(|\bm q|)
	\\
	\label{eq:jLB}
	&\equiv& \xi_{B} B^{\mu},
\end{eqnarray}
where we used the integration by parts with the vanishing surface terms from $F_{\rm W}(0)\rightarrow 0$ and $F_{\rm W}(\infty)\rightarrow 0$. 

Additionally, we should include the electric current of right-handed positrons induced by antineutrinos. This contribution takes the form of Eq.~(\ref{eq:jLB}) by replacing $F_{\rm W}$ with $-\tilde{F}_{\rm W}(|\bm q|)=-F_{\rm W}(|\bm q|)_{\rm n\leftrightarrow \rm p,\, \mu_{\rm e}\rightarrow -\mu_{\rm e}}$, where the minus sign stems from opposite chirality. In fact, the left-handed neutrinos and right-handed antineutrinos move along the same direction as the magnetic field since they are driven by scattered electrons and positrons with both opposite charges and chiralities. Consequently, the total electric current induced by the magnetic field is given by 
\begin{eqnarray}\nonumber
	j_{B,{\rm tot}}^{\mu}
	&=& \frac{\hbar e^2}{4\pi^2} x_0 B^{\mu} \int^{\infty}_0 \frac{{\rm d}|\bm q|}{|\bm q|}\big[F_{\rm W}(|\bm q|) - \tilde{F}_{\rm W}(|\bm q|)\big]
	\\
	&\equiv& \xi^{\rm tot}_{B} B^{\mu}.
\end{eqnarray}
Although the linear growth of $j_{B,{\rm tot}}^{\mu}$ in time is due to the time-independent $f^{(\nu)}(q_0)$, the current is generally enhanced with time provided $F_{\rm W}-\tilde{F}_{\rm W}$ does not flip the sign in time.

We can also evaluate the energy current of electrons driven by neutrino backreaction with magnetic fields via
\begin{eqnarray}\nonumber
	T_{B}^{\mu 0}
	&=&\frac{\hbar e}{8\pi^2} x_0 B^{\mu} \int^{\infty}_0 {\rm d}|\bm q|F_{\rm W}(|\bm q|)\,
		\\
	&\equiv & \kappa_{B} e B^{\mu}.
\end{eqnarray}
Incorporating the similar contribution from positrons, the total energy current is given by 
\begin{eqnarray}\nonumber
	T_{B,{\rm tot}}^{\mu 0}&=&\frac{\hbar e}{8\pi^2} x_0 B^{\mu} \int^{\infty}_0 {\rm d}|\bm q|\big[F_{\rm W}(|\bm q|)+\tilde{F}_{\rm W}(|\bm q|)\big]
	\\
	&\equiv & \kappa_B^{\rm tot} e B^{\mu}.
\end{eqnarray}
Note that the opposite chirality is compensated by the opposite charge unlike the case for $j^{\mu}_{B}$.

We now extract the numerical values for variables involved in $F_{\rm W}$ in the gain region where neutrino absorption dominates over neutrino emission. From Ref.~\cite{Muller:2020ard}, we take the electron fraction $Y_{\rm e} \approx 0.4$, mass density $\rho_{\rm gain} \sim 10^{10}\,{\rm g}\cdot{\rm cm}^{-3}$, and temperature $T \sim 10^{11}\,{\rm K} \approx 8.6\,{\rm MeV}$. Based on charge neutrality, we impose $n_{\rm e}=n_{\rm p} \approx 0.4n_{\rm gain}$ and $n_{\rm n}\approx 0.6n_{\rm gain}$ and approximate $\rho_{\rm gain} \approx M(n_{\rm p}+n_{\rm n})=M n_{\rm gain}$ with the nucleon mass $M \approx 940\,{\rm GeV}$, which yields $\mu_{\rm e} \approx 0.79\,{\rm MeV}$, $\mu_{\rm n} \approx 870\,{\rm MeV}$, and $\mu_{\rm p} \approx 867\,{\rm MeV}$ in equilibrium. 
Adopting the numerical values $\alpha=2.65$ and $\bar{\epsilon}=13.05\,{\rm MeV}$ from Ref.~\cite{Tamborra:2012ac} as an example, we find $\xi_{B} \approx 2.2\,{\rm MeV}$, $\xi^{\rm tot}_{B}\approx -0.5 \,{\rm MeV}$, $\kappa_B\approx 350\,{\rm MeV}^2$, and $\kappa_B^{\rm tot} \approx 780\,{\rm MeV}^2$ for $x_0=0.1\,{\rm s}$. 
The scale of $\xi^{\rm tot}_{B}$ around MeV, despite being a rough estimate, may originate from $\mu_{\rm p}-\mu_{\rm n} \approx -3\,{\rm MeV}$. One can show the overall sign of $\xi^{\rm tot}_{B}$ just depends on the sign of $\mu_{\rm p}-\mu_{\rm n}$ when $\mu_{\rm e}\ll T$. Also, the persistent neutrino emission in the astrophysical timescale compensates the weakness of $G_{\rm F}$, as expected in the scenario of neutrino heating in CCSN. The validity of the linear fluctuation breaks down when $\delta f_{\rm W}^{(\rm e)}\sim 1$ for $x_0 \sim 1\,{\rm s}$, and we may regard the above values of $\xi^{\rm tot}_{B}$ and $\kappa_B^{\rm tot}$ for $x_0=0.1\,{\rm s}$ as approximate upper bounds. 

In light of the results of numerical simulations of the chiral magnetohydrodynamics including the effective CME in Refs.~\cite{Masada:2018swb,Matsumoto:2022lyb} (see also Ref.~\cite{Brandenburg:2017rcb} in the context of the early Universe), we may derive several consequences. First, the effective CME with this $\xi^{\rm tot}_{B}$ generates a magnetic field with a strength $\sim 10^{16}\,{\rm G}$ and magnetic helicity density $\sim -(1$--$10\,{\rm MeV})^3$ via the CPI.
Second, this order of magnitude of $\xi^{\rm tot}_{B}$ is sufficiently large to lead to the inverse cascade of magnetic and fluid kinetic energies. While the previous studies focus on the region within proto-neutron stars, the present result suggests the inverse cascade even in the gain region. This feature should be contrasted with the conventional neutrino radiation hydrodynamics without chiral effects that shows the direct cascade in three spatial dimensions~\cite{Hanke:2011jf,Takiwaki:2013cqa,Radice:2017kmj}.

Finally, based on the momentum conservation, we can also estimate the kick velocity of the proto-neutron star with the core density $\rho_{\rm core}\approx M n_{\rm core}$ and $n_{\rm core} \sim 0.1\,{\rm fm}^{-3}$ due to the effective CME as
\begin{eqnarray}
	v_{\rm kick}\sim \frac{|T_{B,{\rm tot}}^{i 0}|}{\rho_{\rm core}}\approx \left(\frac{eB}{10^{13{\rm -}14}\,\rm G}\right) {\rm km/s}\,.
\end{eqnarray}
It is approximately the same as the upper bound previously obtained in Ref.~\cite{Yamamoto:2021hjs} that assumes neutrinos close to equilibrium (no such an assumption in the present derivation). Inserting the magnitude of the magnetic field due to the CPI above, the resulting kick velocity is $v_{\rm kick}=100$--$1000\,{\rm km/s}$, which is comparable to the observed magnitudes of pulsar kicks~\cite{Lai_2001}.

\sect{Summary}%
In conclusion, we have derived an effective CME triggered by neutrino radiation through the neutrino absorption on nucleons and its inverse process. From there, the electric and energy currents propagating along a magnetic field can be enhanced by persistent neutrino emission in time. Unlike the conventional CME, this effect is purely nonequilibrium and appears even with chirality flipping of electrons. Our findings provide not only a possible mechanism of pulsar kicks but also a strong argument for the inverse cascade of the magnetic and kinetic energies in the gain region of CCSN. These scenarios should be numerically checked by the first-principles calculations of the chiral radiation hydrodynamics for neutrinos. It would also be interesting to apply the present framework to other systems, such as neutron star mergers and cosmology in the early Universe.

\acknowledgments
\sect{Acknowledgments}%
The authors thank B.~M\"uller, T.~Takiwaki, and M.-R.~Wu for fruitful discussions.
N.~Y. is supported by the Keio Institute of Pure and Applied Sciences (KiPAS) project at Keio University and JSPS KAKENHI Grant No.~JP19K03852 and No.~JP22H01216.
D.-L.~Y. is supported by National Science and Technology Council (Taiwan) under Grant No.~MOST 110-2112-M-001-070-MY3.

\bibliography{Effective_CME_neutrinos_v2.bbl}
\end{document}